\def\be{\begin{equation}}
\def\ee{\end{equation}}
\def\bea{\begin{eqnarray}}
\def\eea{\end{eqnarray}}
\def\a{\alpha}

\def\t{\tau}

\documentclass[12pt]{article}
\begin{document}
\title{Dimers on a simple-quartic net with a vacancy}
\author{W.-J. Tzeng$^1$ and F. Y. Wu$^2$ \\ $^1$Department of Physics, Tamkang University \\ Tamsui, Taipei, Taiwan 251, R. O. C.\\ $^2$Department of Physics, Northeastern University\\Boston, Massachusetts 02115}

\date{}
\maketitle

\begin{abstract}

A seminal milestone in lattice statistics is the exact solution of the
enumeration of dimers on a simple-quartic net obtained by Fisher,
Kasteleyn, and Temperley (FKT) in 1961.
An outstanding related and yet unsolved problem is the enumeration of dimers
on a net with vacant sites.
Here we consider this vacant-site problem with a single vacancy
occurring at certain specific sites on the boundary of a simple-quartic net.
First, using a bijection between dimer and spanning tree configurations
due to Temperley, Kenyon, Propp, and Wilson, we establish that
the dimer generating function is independent of the location of the vacancy,
and deduce a closed-form expression for the generating function.
We next carry out finite-size analyses of this solution
as well as that of the FKT solution.
Our analyses lead to a logarithmic correction term in the
large-size expansion for the vacancy problem with free boundary conditions.
A concrete example exhibiting this difference is given.
We also find the central charge $c=-2$ in the language of conformal field theory
for the vacancy problem,as versus the value $c=1$ when there is no vacancy.

\end{abstract}

\newpage

\section{Introduction}

The problem of enumerating close-packed dimers on a finite simple-quartic net was solved by Temperley and Fisher \cite{temp,fisher61} and by Kasteleyn \cite{kas} in 1961.
An outstanding related but yet unsolved problem is the enumeration of dimers
on a net with vacant sites \cite{kenyon}.
Here, we consider this vacancy problem when a single vacant site occurs
on the boundary.
\medskip

The difficulty associated with the vacancy problem is that, while the determinant whose
square root yieids the dimer generating function can be written down using the Kasteleyn
formulation \cite{kas}, its evaluation is difficult.
In 1974 Temperley \cite{temperley1} reported an intriguing bijection
relating close-packed dimer coverings with spanning tree configurations
on two related lattices.
This offers an alternate approach to the vacancy problem since spanning trees
can be enumerated by standard means.
The Temperley bijection has been of renewed recent interest
and extended to graphs with certain
weighted and/or directional edges \cite{kpw}.
Here, we use an extension of the Temperley bijection due to Kenyon, Propp, and
Wilson \cite{kpw} to study the vacancy dimer problem.
\medskip

Our first result is that, for a simple-quartic net with free boundaries and one fixed vacant site located at certain specific sites on the boundary,
the dimer generating function is independent of the position of the vacancy.
The exact generating function for close-packed dimers
on this net is then deduced from that of the spanning trees.
In view of the connection with the conformal field theory \cite{cardy} and current interests in finite-size analyses for two-dimensional lattice models \cite{Ferdinand67,LuWu99,csm},
we next carry out finite-size analyses for the weighted spanning tree
solution as well as that of the FKT solution.
It is found that a logarithmic correction term arises in the large-size expansion 
in the case of the vacancy problem with free boundaries, 
a term which is absent in the expansion of the FKT solution.
A concrete example demonstrating this difference is given.
We also find that the occurrence of a vacancy yields a new central charge $c=-2$
in the language of the conformal field theory.
\medskip

The organization of this paper is as follows: 
To make the paper self-contained, we restate and establish in section 2
the bijection due to Temperley \cite{temperley1} between spanning tree and dimer configurations, as well as an extended version of the bijection due to
Kenyon, Propp, and Wilson \cite{kpw}.
This extended Temperley bijection permits us to establish in section 3
the independence of the dimer generating function on the location
of the vacancy when it occurs at specific boundary sites.
The explicit expression of the generating function for a simple-quartic
net is then obtained.
In section 4 we carry out finite-size analyses for weighted spanning trees
and the FKT dimer solution. 
Specializing the results to dimer enumerations, we find a logarithmic correction term which is unique to the vacancy problem with free boundaries.
Discussions and a summary of our findings are given in section 5.

\section{The extended Temperley bijection}

For definiteness we consider a simple-quartic net (lattice) with free boundaries,
although much of the results of this section also hold for more general planar graphs \cite{kenyon,kpw}.
\medskip

First we restate the Temperley bijection.
Starting from a $L_1\times L_2$ simple-quartic net $G$ with free boundaries,
one constructs a dimer lattice $G_D$ by i) adding a new site at the midpoint
of each edge of $G$,
ii) inserting in each internal face of $G$ a new site connected to
the midpoints of the 4 edges of $G$ surrounding it, and
iii) removing one corner site of the resulting lattice
and its incident edges on $G_D$.
Thus, $G_D$ has a total of $(2L_1-1)(2L_2-1)-1$ sites consisting of
the original $L_1L_2-1$ sites of $G$, which we call the {\it odd} sites,
and the remaining $(2L_1-1)(2L_2-1)-L_1L_2$ new sites,
which we call the {\it even} sites.
An example of this construction for $L_1=L_2=3$ is shown in Figs. 1(a) and 1(b).
\medskip

A spanning tree is a collection of connected edges of $G$ which does not form closed
circuits and covers all sites. 
Then we have the
\medskip

{\it Temperley bijection: There exists a one-one correspondence between
spanning tree configurations on $G$ and dimer configurations on $G_D$.}
\medskip

To see that the bijection holds, one observes that to each spanning tree configuration
on $G$, one can construct a unique dimer configuration on $G_D$ by
first laying a dimer along each tree edge, starting from the edge(s)
covering the corner site of $G_D$ which has (have) been removed,
and proceed along the spanning tree edges in an obvious fashion.
After laying dimers along all tree edges, the remaining sites of $G_D$
can then be covered by dimers in a unique way \cite{temperley1}.
Conversely, starting from each dimer configuration on $G_D$,
one constructs a unique tree configuration on $G$ by drawing bonds
(tree edges) along dimers originating from all odd sites.
These bonds cannot form close circuits, since otherwise
they would have enclosed an odd number of sites of $G_D$ which is not permitted
in close-packed dimer configurations.
This process leads to a unique tree configuration on $G$.
This completes the proof.
An example of the Temperley bijection is shown in Figs.2(a) and 2(b).
\medskip

Kenyon, Propp, and Wilson \cite{kpw} have shown that
the Temperley bijection holds more generally for graphs with certain
weighted and/or directed edges. For our purposes, however, we shall
confine ourselves to the original Temperley bijection as
stated in the above, with  the step iii)replaced by one of
removing an odd site on the boundary together with its incident edges.
An example of constructing such a $G_D$ is shown in Fig.1(c).
The proof of the bijection between tree configurations on $G$
and dimer coverings on $G_D$goes through as before,
and we are led to the
\medskip

{\it Extended Temperley bijection (Temperley-Kenyon-Propp-Wilson):
There existsa one-one correspondence between spanning trees on $G$
and dimer coverings on any $G_D$ constructed from $G$
by removing any boundary odd site and its incident edges in step
iii) of the construction described in the above.}
\medskip
\noindent

An example of such a bijection is shown in Figs.2(a) and 2(c).
\medskip

{\it Remark}:
The extended Temperley bijection does not hold for dimer lattices $G_D$ containing
an interior vacancy.
In that case while each spanning tree can still be mapped
into a unique dimer configuration as before,
there exist dimer configurations which cannot be mapped into spanning trees.
These are dimer coverings with no dimers laying on any of the 4 edges
of $G$ incident to the defect site.

\section{Dimer lattice with a vacant boundary site}
\subsection{Dimer generating function}

The dimer generating function for $G_D$ is
\begin{equation}
Z(G_D;x_1,x_2) =
\sum_{\rm dimer \>config.}x_1^{n_1}x_2^{n_2} \label{dimer}
\end{equation}
where the summation is taken over all dimer covering configurations,
$x_1$ and $x_2$ are, respectively, the weights of horizontal and vertical dimers.,
and $n_1$ and $n_2$ are, respectively, the number of horizontal and vertical dimers.
Clearly, we have
\begin{equation}
Z(G_D;1,1) ={\rm the\>\>number\>\>of\>\>
dimer\>\> configurations\>\> on\>\>} G_D.\label{dimernumber}
\end{equation}
\medskip

Consider two different dimer lattices $G_D$ and $G_D'$ obtained from $G$
as described in the above, namely, by removing different boundary odd sites.
Then we have the following equivalence:
\medskip

{\it Proposition 3.1.1}:
\begin{equation}
Z(G_D;x_1,x_2) = Z(G_D';x_1,x_2) \label{theorem}
\end{equation}
{\it for any $G_D$ and $G_D'$}.
\medskip

{\it Proof}:
The extended Temperley bijection dictates that
there is a one-one correspondence between spanning tree configurations
on $G$ and dimer configurations on {\it any} $G_D$.
It follows that there is a bijection between dimer coverings on $G_D$
and $G_D'$, and that the summation in (\ref{dimer}) on $G_D$ can
be considered as taken over all spanning tree configurations on $G$.
\medskip

For each spanning tree configuration $T$ of $G$,
the dimer weight in the summand in (\ref{dimer}) consists of two factors,
\begin{equation}
x_1^{n_1}x_2^{n_2} = W_o(T;x_1,x_2)W_e(T;x_1,x_2),\label{prod}
\end{equation}
where $W_o$ is the product of the weights of those dimers
originating from odd sites, and $W_e$ is the product of the weights
of those dimers covering two even sites. 
For the two dimer coverings of $G_D$ and $G_D'$ corresponding to the same$T$, 
their $W_e$ factors are the same by definition. 
Their $W_o$ factors are also the same since, 
even though the respective dimer positions may be shifted, 
they lay along the same spanning tree edges hence carry the same weights. 
It follows that summations on the lhs and rhs of (\ref{theorem}) are identical term by term, 
and the proposition is proved. {\hfill{$\bullet$}}
\medskip

{\it Remark}: Proposition 3.1.1 holds more generally for arbitrary planar $G$ and
its related $G_D$.
Since the overall dimer weight can always be factorized into the product $W_oW_e$ 
as in (\ref{prod}), the proof of the proposition goes through as presented.
\medskip

We next consider the generating function of {\it weighted} spanning trees for
the $L_1\times L_2$ simple-quartic $G$.
Assign weights $x_1$ and $x_2$, respectively,
to edges in the horizontal and vertical direction.
Then, the weighted spanning tree generating function is
\begin{equation}
T(G;x_1,x_2) = \sum_{T}x_1^{n_1}x_2^{n_2} \label{tree}
\end{equation}
where the summation is taken over all spanning tree configurations $T$ on $G$ and, 
as in (\ref{dimer}), $n_1$ and $n_2$ are the numbers of edges in the spanning tree 
in the respective directions.
Particularly, we have
\begin{equation}
T(G;1,1) ={\rm the\>\>number\>\>of\>\>
spanning\>\>tree\>\> configurations\>\> on\>\>} G.\label{treenumber}
\end{equation}
{}From the extended Temperley bijection, it is clear that we can also write (\ref{tree}) 
as
\begin{equation}
T(G;x_1,x_2) = \sum_{T} W_o(T;x_1,x_2) \label{tree1}
\end{equation}
where $W_o$ is the factor in (\ref{prod}) for the dimer covering on any $G_D$.
It is seen from (\ref{tree1}) that if all dimers covering even sites of $G_D$
have the weight $1$, namely $W_e=1$,
then the dimer generating function is simply $ T(G;x_1,x_2)$.
\medskip

More generally for a simple-quartic $G$ of size $L_1\times L_2$ and the related $G_D$,
we have the equivalence:
\medskip

{\it Proposition 3.1.2}:
\begin{equation}
Z(G_D;x_1,x_2) = x_1^{L_1(L_2-1)}x_2^{L_2(L_1-1)} T\bigg(G;{{x_1}
\over {x_2}},{{x_2}\over {x_1}}\bigg). \label{theorem1}
\end{equation}
\medskip

{\it Proof}: From the construction of $G_D$ we note that there are a total of 
$L_2(L_1-1)$ even sites located at midpoints of horizontal edges of $G$.
We call these the $H$ sites. 
Similarly, there are $L_1(L_2-1)$ even sites of $G_D$ located 
at midpoints of vertical edges of $G$, which we call the $V$ sites.
\medskip

The $H$ and $V$ sites can be covered by either horizontal or vertical dimers.
Let $N_{Hh}$ ($N_{Hv}$) be the number of horizontal (vertical) dimers covering
the $H$ sites. 
Then we have
\begin{equation}
N_{Hh} + N_{Hv} = L_2(L_1-1).
\label{sum1}
\end{equation}
Likewise, we have
\begin{equation}
N_{Vh} + N_{Vv} = L_1(L_2-1),
\label{sum2}
\end{equation}
where $N_{Vh}$ ($N_{Vv}$) is the number of horizontal (vertical)dimers covering the $V$ sites. 
In these notation, we can rewrite the spanning tree and dimer generating functions as
\begin{eqnarray}
T(G;x,y) &=& \sum_T x^{N_{Hh}} y^{N_{Vv}} \nonumber \\
Z(G_D;x_1,x_2) &=& \sum_T {x_1}^{N_{Hh}+N_{Vh}} {x_2}^{N_{Hv}+N_{Vv}}, \label{gene1}
\end{eqnarray}
where we have used the one-one correspondence between spanning tree 
configurations $T$ and dimer configurations. 
The Proposition 3.1.2 now follows after eliminating $N_{Vh}$ and $N_{Hv}$ in
(\ref{gene1}) using (\ref{sum1}) and (\ref{sum2}). {\hfill{$\bullet$}}

\subsection{Dimer enumerations}

For a simple-quartic $G$ of size $L_1\times L_2$ with free boundaries,
the generating function (\ref{tree1}) for weighted spanning trees
has been evaluated \cite{TzWu00} and is given by
\begin{eqnarray}
&&T(G;x_1,x_2)=
\frac{1}{L_1L_2}\prod_{m=0}^{L_1-1} \prod_{n=0}^{L_2-1}
\Bigg[2\ x_1\Bigg(1-\cos\frac{m\pi}{L_1}\Bigg)
+2\ x_2\Bigg(1-\cos\frac{n\pi}{L_2}\Bigg)\Bigg],\nonumber \\
&& \hskip 6cm ~(m,n)\not=(0,0). \label{tree2}
\end{eqnarray}
Now the dimer lattice $G_D$ is of size $M\times N$ with a boundary vacancy, where
\begin{equation}
M=2L_1-1, \hskip 1cm N=2L_2-1.\label{mn}
\end{equation}
In terms of the dimer lattice sizes $M,N$, we thus have,
after using Proposition 3.1.2 and (\ref{tree2}),
\begin{eqnarray}
&&Z_{\{M\times N-1\}}(x_1,x_2)= 
x_1^{(M-1)/2}x_2^{(N-1)/2}\nonumber\\ 
&&\times\ \prod_{m=1}^{\frac{M-1}{2}}\prod_{n=1}^{\frac{N-1}{2}} 
\left[4{x_1}^2\cos^2\frac{m\pi}{M+1}+4{x_2}^2\cos^2\frac{n\pi}{N+1}\right].
\label{dimergd}
\end{eqnarray}
Here, we must have $MN=$ odd to admit dimer coverings. 
The subscript $\{M\times N-1\}$ in (\ref{dimergd}) reminds us that the enumeration is 
for an $M\times N$ net with one boundary odd site removed.
Note that the factor $m=n=0$ is excluded in the product in (\ref{dimergd}).
This expression is to be compared with the enumeration of dimers
on an $M\times N$ simple-quartic net without vacancies.
For $M$ and $N$ both even, for example, the expression is \cite{kas}
\begin{equation}
Z_{\{M,N\}}(x_1,x_2) =\prod_{m=1}^{M/2}\prod_{n=1}^{N/2}
\left[4x_1^2\cos^2\left(\frac{m\pi}{M+1}\right)
+4x_2^2\cos^2\left(\frac{n\pi}{N+1}\right)\right].\label{dimer2}
\end{equation}

\section{Finite-size analyses}

Finite-size expansions of physical quantities associated with two-dimensional lattice
models have been of current interest both in physics \cite{cardy,LuWu99,csm} and in
mathematics \cite{kenyon}.
For the dimer problem Kenyon \cite{kenyon} has recently deduced very general
results on the leading terms of the asymptotic expansion of the dimer enumeration for
rectilinear lattices with free boundaries of any shape.
Alternately, one can obtain expansions for regular lattices, in principle to all orders,
by analyzing known exact expressions\cite{Ferdinand67,LuWu99}. 
This is the approach we now use.

\subsection{Spanning tree generating function}

Consider first the generating function (\ref{tree2}) for the fully weighted spanning trees.
For large $L_1$ and $L_2$, we expect to have
\begin{equation}
{1\over {L_1L_2}} \ln T(G;x_1,x_2)= f_{\rm bulk}(x_1,x_2) + f_c(x_1,x_2)
\label{expansion}
\end{equation}
where $f_{\rm bulk}$ is the per-site bulk free energy and $f_c$ is
the correction containing terms of the order of $L_1^{-1}$, $L_2^{-1}$ and higher.
Using (\ref{tree2}), we find the bulk free energy
\begin{eqnarray}
f_{\rm bulk}(x_1,x_2) &\equiv& \lim_{L_1,L_2 \to \infty} {1\over {L_1L_2}}
\ln T(G;x_1,x_2) \nonumber \\
&=& {1\over {\pi^2}} \int_0^\pi d\theta \int_0^\pi d\phi
\ln \Big[2\ x_1(1-\cos \theta) +2\ x_2(1-\cos \phi) \Big] \nonumber \\
&=& \frac 4 {\pi} \int_0^{\pi/2}d\phi
\ln\Big( \sqrt{x_1+x_2 \sin^2\phi}
+\sqrt{x_2} \sin\phi \Big),
\label{free}
\end{eqnarray}
where the last line is obtained by carrying out the $\theta$ integration.
The computation of correction terms $f_c$ for products of the form
of (\ref{tree2}) is standard \cite{fisher61,Ferdinand67,barber}.
Particularly, one has
\be
f_{\rm bulk}(1,1) = \frac 4 \pi G
\ee
where $G$ is the Catalan constant given by
\be
G=1 -3^{-2} +5^{-2} -7^{-2}
+\cdots =0.915\ 965\ 594\ldots.
\ee
\medskip

To compute (\ref{expansion}) we proceed as follows.
Take out a factor $x_1$ from each of the $L_1L_2-1$ factors in(\ref{tree2}) and 
split the product into 3 parts to take care of the exclusion of the $m=n=0$ factor.

We have
\begin{equation}
T(G;x_1,x_2) = \Big(L_1L_2\Big)^{-1} x_1^{L_1L_2-1}\ \big(T_0T_1T_2\big)
\end{equation}
where 
\begin{equation}
T_0 =\prod_{m=1}^{L_1-1}\prod_{n=1}^{L_2-1} F(m,n), \quad
T_1 =\prod_{m=1}^{L_1-1} F(m,0), \quad T_2 =\prod_{n=1}^{L_2-1} F(0,n)
\end{equation}
with
\begin{equation}
F(m,n) =2\Bigg(1-\cos\frac{m\pi}{L_1}\Bigg)+2\tau \Bigg(1-\cos\frac{n\pi}{L_2}\Bigg)
\end{equation}
and
\be
\t=x_2/x_1.
\ee
Using the identity \cite{Grad94}
\begin{equation}
\prod_{n=1}^{N-1} \Bigg[2 \cosh 2\theta - 2 \cos {{n\pi}\over N} \Bigg] =
{{\sinh 2N \theta} \over {\sinh 2\theta}}, \label{id}
\end{equation}
and its $\theta \to 0$ limit,
\be
\prod_{n=1}^{N-1} \Bigg[2 - 2 \cos {{n\pi}\over N} \Bigg] = N,
\ee
we find
\begin{equation}
T_0=\prod_{n=1}^{L_2-1} {{\sinh (2L_1\theta_n)} \over {\sinh 2\theta_n}},
\hskip 1cm
T_1 = L_1, \hskip 1cm T_2= \tau ^{L_2-1}L_2, \label{t012}
\end{equation}
with $\theta_n$ given by
\begin{equation}
\cosh 2\theta_n = 1+ \tau \Bigg(1-\cos{{n\pi}\over {L_2}}\Bigg),
\end{equation}
or, explicitly,
\bea
\theta_n =F\Big( \frac{n\pi}{2L_2}\Big) &\equiv&
\big[\cosh^{-1}\big(1+2a_n^2\big)\big]\big/2 \nonumber \\
&=& \sinh^{-1} a_n \nonumber \\
&=& \ln \Big(a_n +\sqrt{1+a_n^2}\Big), \label{z}
\eea
where
\be
a_n=\sqrt{\t} \sin \frac{n\pi}{2L_2}.
\ee
Substituting (\ref{t012}) into (\ref{tree2}), we thus obtain
\be
T(G;x_1,x_2)= x_1^{L_1L_2-1} \tau ^{L_2-1} \prod_{n=1}^{L_2-1}
{{\sinh (2L_1\theta_n)} \over {\sinh 2\theta_n}}. \label{tree3}
\ee
The product in the denominator in (\ref{tree3}) can again be evaluated
using (\ref{id}) as
\bea
\prod_{n=1}^{L_2-1}\sinh^22 \theta_n
&=& \prod_{n=1}^{L_2-1}(\cosh 2\theta_n-1)\cdot
\prod_{n=1}^{L_2-1}(\cosh 2\theta_n+1) \nonumber \\
&=& \prod_{n=1}^{L_2-1}\Big[\t\Big(1-\cos \frac{n\pi}{L_2}\Big)\Big]\cdot
\prod_{n=1}^{L_2-1}\Bigg[2+\t-\t\cos \frac{n\pi}{L_2}\Bigg] \nonumber \\
&=& {L_2}\Bigg({\t\over 2}\Bigg)^{2(L_2-1)}
\Bigg({{\sinh 2L_2\a} \over {\sinh 2\a}}\Bigg), \label{p}
\eea
where $\a$ is given by
\be
\cosh 2\ \a = 1+2\t^{-1}
\ee
or $\ \sinh\a=1/\sqrt{\t}$, or explicitly,
\be
\a = \ln \Big(\sqrt{\t^{-1}}+\sqrt{1+\t^{-1}}\Big).
\ee
Combining these results, we obtain from (\ref{tree3}) the expression
\be
T(G;x_1,x_2)= x_1^{L_1L_2-1}
\Bigg({{\sinh 2\a} \over {L_2\sinh 2L_2\a}}\Bigg)^{1/2} \ \prod_{n=1}^{L_2-1}
\Big[2\sinh\big(2L_1\theta_n\big)\Big]. \label{tree4}
\ee
Taking the logarithm, we obtain
\bea
&&\ln T(G;x_1,x_2)=(L_1L_2-1) \ln x_1 +2L_1 \sum_{n=1}^{L_2-1}
{\theta_n} + \sum_{n=1}^{L_2-1} \ln \big(1-e^{-4 L_1\theta_n}\Big)
\nonumber\\
&& \hskip .5cm
-L_2\ \a -\frac 1 2 \ln\Big(1-e^{-4L_2\a}\Big)-\frac 1 2 \ln L_2
+\frac 1 2 \ln(\sinh 2\a) .
\label{tree5}
\eea
\medskip

For large $L_1$ and $L_2$ with the ratio $L_1/L_2$ finite,
the first two terms in (\ref{tree5}) contribute to the bulk free energy
$f_{\rm bulk}(x_1,x_2)$ given in (\ref{free}).
To carry out the summations in (\ref{tree5}),
we use the Euler-MacLaurin summation formula given by
\begin{eqnarray}
\sum_{n=1}^Nf(a+n\delta)&=&\frac{1}{\delta}\int_a^{a+N\delta}f(x)dx
+\frac{1}{2}\left[f(a+N\delta)-f(a)\right]\nonumber\\
&&+\frac{\delta}{12}\left[f'(a+N\delta)-f'(a)\right]+O(\delta^3).\label{euler}
\end{eqnarray}
With $a=0$, $N=L_2$, $\delta=\pi/2L_2$, and $f(x)=F(x)$ defined in (\ref{z}),
one has
\bea
\sum_{n=1}^{L_2-1}\theta_n &=& \sum_{n=1}^{L_2}\theta_n - \theta_{L_2} \nonumber \\
&=&\frac{L_2}{2}\Big[f_{\rm bulk}(x_1,x_2) - \ln x_1\Big]
-\frac{1}{2}\ln(\sqrt{1+\tau}+\sqrt{\tau}) \nonumber \\
&& \hskip .5cm - \frac{\pi \sqrt{\tau}}{24L_2} +O(L_2^{-3}).
\eea
For the second summation in (\ref{tree5}),
we follow the manipulation in \cite{Ferdinand67} to write
\be
\sum_{n=1}^{L_2-1} \left[1-e^{-4L_1\theta_n}\right] =\sum_{n=1}^{\infty}
\left(1-e^{-2n {L_1}\pi\sqrt{\tau}/L_2}\right)+O(L_2^{-2+\epsilon})
\ee
for some $0<\epsilon<2$, with $\epsilon\to 0$ when $L_2\to\infty$.
Putting the results together, we find the finite-size correction
\be
f_c(x_1,x_2) = \frac {c_1(x_1,x_2)} {L_1} + \frac {c_2(x_1,x_2)} {L_2}
+ \frac {c_3(x_1,x_2)} {L_1L_2} +o\Big(\frac { 1} {L_1L_2}\Big),
\label{finitetree} \ee where \bea c_1 (x_1,x_2)&=& -\ln \Big(\sqrt{\tau^{-1}} 
+\sqrt {1+\t^{-1}}\Big) \nonumber \\ c_2
(x_1,x_2)&=& -\ln \Big(\sqrt{\tau} +\sqrt {1+\t}\Big) \nonumber \\
c_3(x_1,x_2)&=& - \frac 1 2 \ln L_2 +\ln 2+\frac{1}{4}\ln\frac{(x_1+x_2)}{x_1^3x_2^2}
-\frac{\pi\sqrt{\tau}L_1}{12L_2} \nonumber \\ 
&& \hskip 1cm
+\sum_{n=1}^\infty\ln \left(1-e^{-2n\pi\sqrt{\tau}L_1/L_2}\right).
\label{c3f} 
\eea
Particularly, for $x_1=x_2=1$, the expression $f_c(1,1)$ given by (\ref{finitetree}) 
reduces to the one given in \cite{kenyon} and \cite{dd}.
\medskip

Despite its appearance, the expression for $c_3(x_1,x_2)$ is actually symmetric
in $\{x_1,L_1\} \leftrightarrow \{x_2,L_2\}$, a fact can be seen from the identity
\begin{eqnarray}
&&\sum_{m=1}^\infty\ln
\left(1-e^{-2m\pi L_2/\sqrt{\tau}L_1 }\right)-\frac{1}{2}\ln L_1
-\frac{1}{4}\ln x_2-\frac{\pi L_2}{12\sqrt{\tau}L_1}\nonumber\\
&=&\sum_{n=1}^\infty\ln \left(1-e^{-2n\pi\sqrt{\tau}L_1/L_2}\right)-\frac{1}{2}\ln L_2
-\frac{1}{4}\ln x_1-\frac{\pi\sqrt{\tau}L_1}{12L_2}. \label{transf}
\end{eqnarray}
Introducing the Jacobi theta function
\begin{equation}
\vartheta_1(\phi,q)= 2q^{1/4}\sin\phi\prod_{n=1}^\infty
(1-q^{2n})(1-2q^{2n}\cos 2\phi+q^{4n}),
\end{equation}
and the identity \cite{WG89}
\be
\prod_{n=1}^\infty
\left(1-q^{-2n}\right)
=\left[\frac{\vartheta'_1(0,q)}{2q^{1/4}}\right]^{1/3}
\ee
where $\vartheta'_1$ is the derivative of $\vartheta_1$
with respect to $\phi$,
then we have also
\be
c_3(x_1,x_2) =- \frac 1 2 \ln L_2
+ \frac{1}{4}\ln\frac{(x_1+x_2)}{x_1^3x_2^2}
+ \frac{1}{3} \ln \Big[ 4 \vartheta'_1(0, q) \Big]
\ee
where
\be
q = e^{-L_1\sqrt{\tau}\pi/L_2}.
\label{q}
\ee
The identity (\ref{transf}) follows from the Jacobi transformation \cite{WG89}
\be
\vartheta'_1(0,e^{- \pi v})=v^{-3/2}\vartheta'_1(0,e^{-\pi/v}), \quad v>0.
\ee
Now, the theta function $\vartheta'_1$ is finite for $L_1/L_2$ finite,\footnote{This 
also follows from the fact that the series
$\sum_{n=1}^\infty\ln(1-u^n) =-\sum_{n=1}^\infty\sum_{m=1}^\infty{u^{mn}}/m
=-\sum_{m=1}^\infty{u^m}/{m(1-u^m)}$ converges for all $0\leq u <1$.}
it follows that the leading behavior of $c_3$ is $\ln L_2 \>\>(\sim \ln L_1)$.
\medskip
In conformal field theory \cite {cardy} one needs to compute the limits
\begin{eqnarray}
\frac 1 {L_1} \lim_{L_2\to\infty}\frac{\ln T({\bf Z}_2)}{L_2} &=&f_{\rm bulk}
+\frac{c_1}{L_1} +\frac{\Delta_1}{L_1^2} +o(L_1^{-2});\label{central1}\\
\frac 1 {L_2} \lim_{L_1\to\infty}\frac{\ln T({\bf Z}_2)}{L_1} &=&f_{\rm bulk}
+\frac{c_2}{L_2} +\frac{\Delta_2}{L_2^2} +o(L_2^{-2}).
\label{central2}
\end{eqnarray}
Using (\ref{finitetree}), we find
\be
\Delta_1 = -\frac \pi {12 \sqrt \t}, \hskip 1cm
\Delta_2 = -\frac {\pi \sqrt \t} {12}.
\ee
This yields a central charge $c=-2$ upon taking $x_1=x_2=1$ ($\tau = 1$).

\subsection{Dimer enumerations}

We are now in a position to analyze the finite-size corrections
of the two dimer enumerations (\ref{dimergd}) and (\ref{dimer2}).
Although the expressions refer to two dimer lattices with different geometry, 
one for an $M\times N -1$ lattice with a vacancy and
$MN=$ odd, and one for an $M\times N$ lattice with $MN=$ even,
a comparison can still be meaningful if both expansions are expressed in terms of
lattice sizes $M$ and $N$.
%
\medskip

{\it (a) Close-packed dimers}:
For close-packed dimers on an $M\times N$ net with $MN=$ even, we have carried out
the analysis for the expression (\ref{dimer2}) along the lines outlined in the above,
and obtained the result (which can also be extracted from discussions
in \cite{Ferdinand67,LuWu99})
\be
\ln Z_{\{M\times N\}}(x_1,x_2)= (MN+1) {\bar f}_{\rm bulk} + N{{\bar c}_1}
+ M{{\bar c}_2} + {\bar c}_3 + o( 1 ) , \label{dimer6} 
\ee
where
\bea
{\bar f}_{\rm bulk}(x_1,x_2) &=& \frac{1}{4} f_{\rm bulk}(x_1^2,x_2^2), \nonumber\\ 
{\bar c}_1(x_1,x_2)&=&{\bar f}_{\rm bulk}(x_1,x_2)
-\frac{1}{2}\ln\left(x_1+\sqrt{x_1^2+x_2^2}\right) \nonumber \\ 
{\bar c}_2(x_1,x_2)&=&{\bar f}_{\rm bulk}(x_1,x_2)
-\frac{1}{2}\ln\left(x_2+\sqrt{x_1^2+x_2^2}\right) \nonumber \\ 
{\bar c}_3(x_1,x_2) &=&\frac{1}{2}\ln 2
-\frac{1}{2}\ln\left(x_1+\sqrt{x_1^2+x_2^2}\right)
-\frac{1}{2}\ln\left(x_2+\sqrt{x_1^2+x_2^2}\right)\nonumber\\ 
&&+\frac{1}{4}\ln(x_1^2+x_2^2)+\frac{\pi Mx_2}{24Nx_1} +\sum_{n=1}^\infty\ln
\left(1+e^{-(2n-1)\pi Mx_2/Nx_1}\right).\nonumber \\
\label{dimercorrection1}
\eea 
Expressions of ${\bar c}_1(x_1,x_2)$ and ${\bar c}_2(x_1,x_2)$ reduce to those found 
by Kenyon\cite{kenyon} when $x_1=x_2=1$. 
In the language of the conformal field theory\cite{cardy}, 
the term $\pi Mx_2/24 Nx_1$ in ${\bar c}_3$ yields the central charge $c=1$
upon taken $M=N$ and $x_1=x_2$, the accepted value for dimer and Ising systems.
\medskip

Again, the expression (\ref{dimercorrection1}) for ${\bar c}_3$ is
symmetric in $\{x_1,M\} \leftrightarrow \{x_2,N\}$,
a fact can be seen from the identity
\begin{equation}
\frac{\pi Mx_2}{24Nx_1}+\sum_{n=1}^\infty\ln \left(1+e^{-(2n-1)\pi
Mx_2/Nx_1}\right) =\frac{\pi Nx_1}{24Mx_2}+\sum_{m=1}^\infty\ln
\left(1+e^{-(2m-1)\pi Nx_1/Mx_2}\right). \label{exchange2}
\end{equation}
The series $\sum_{n=1}^\infty\ln(1+u^{2n-1})$ converges,\footnote{This also follows 
from the fact that the series
$\sum_{n=1}^\infty\ln(1+u^{2n-1}) =-\sum_{m=1}^\infty{(-u)^m}/{m(1-u^{2m})}$ 
converges for all $0\leq u <1$.}
so ${\bar c}_3$ does not diverge for large $M,N$.
\medskip

The expression for ${\bar c}_3$ can also be written as
\begin{eqnarray}
{\bar c}_3(x_1,x_2) &=& \frac{2}{3}\ln 2-\frac{1}{2}\ln(x_1+\sqrt{x_1^2+x_2^2})
-\frac{1}{2}\ln(x_2+\sqrt{x_1^2+x_2^2})\nonumber\\
&&+\frac{1}{4}\ln(x_1^2+x_2^2) +\frac 1 2 \ln\vartheta_3(0,q)
-\frac 1 6 \ln \vartheta'_1(0,q),
\end{eqnarray}
where $q=e^{-\pi Mx_2/Nx_1}$ and the theta function $\vartheta_3$ is given by \cite{WG89}
\begin{equation}
\vartheta_3(\phi,q)=\prod_{n=1}^\infty (1-q^{2n})(1+2q^{2n-1}\cos2\phi+q^{4n-2}).
\end{equation}
Then, the identity (\ref{exchange2}) is a consequence of
the Jacobi transformations (\ref{transf}) and
\be
\vartheta_3(0,e^{-\pi/v})=v^{1/2}\vartheta_3(0,e^{-\pi v}),\quad v>0.
\ee
\medskip

{\it (b) Close-packed dimers with a boundary vacancy}:
For the simple-quartic net $M\times N -1$ with a boundary vacancy and $MN=$ odd,
one uses 
\be
L_1=(M+1)/2, \hskip 1cm L_2=(N+1)/2, 
\ee 
and expand (\ref{finitetree}) for large $M$ and $N$. 
After some algebra, we find
\bea 
&& \ln Z_{\{M\times N -1\}}(x_1,x_2)\nonumber\\ 
&=&(L_1L_2-L_1)\ln x_1 +(L_1L_2-L_2)\ln x_2 
+\ln T\Bigg(G;\frac{x_1}{x_2},\frac{x_2}{x_1}\Bigg) \nonumber\\ 
&=&(L_1L_2-L_1)\ln x_1 +(L_1L_2-L_2)\ln x_2 + L_1L_2 
\Bigg[f_{\rm bulk}\Bigg(\frac{x_1}{x_2},\frac{x_2}{x_1}\Bigg) +
f_c\Bigg(\frac{x_1}{x_2},\frac{x_2}{x_1}\Bigg)\Bigg] \nonumber \\
&=&( MN+1){\bar f}_{\rm bulk} + N{{\bar c}_1}
+ M{{\bar c}_2} +{\bar c}_3' +o(1), \label{dimer7} 
\eea 
where
${\bar c}_1$ and ${\bar c}_2$ are given in (\ref{dimercorrection1}), and
\begin{eqnarray}
{\bar c}_3'(x_1,x_2) &=&\frac{3}{2}\ln -\frac{1}{2}\ln\Big(x_1+\sqrt{x_1^2+x_2^2}\Big)
-\frac{1}{2}\ln\Big(x_2+\sqrt{x_1^2+x_2^2}\Big)
+\frac{1}{4}\ln\Bigg(1+\frac{x_2^2}{x_1^2}\Bigg) \nonumber\\
&&-\frac{1}{2}\ln N -\frac{\pi Mx_2}{12Nx_1} +\sum_{n=1}^\infty\ln
\left(1-e^{-2n\pi Mx_2/Nx_1}\right). \label{dimercorrection2}
\end{eqnarray}
Comparing ${\bar c}_3'$ with ${\bar c}_3$ given by (\ref{dimercorrection1}), 
we see that for large $M,N$ the deletion of a boundary site introduces 
a logarithmic correction term $-\ln\sqrt N$ in ${\bar c}_3'$.
Furthermore, upon taking $M=N$ the term $-\pi Mx_2/12 Nx_1$ in ${\bar c}_3'$ yields a
central charge $c=-2$, which is different from that of the dimer system without vacancies.
\medskip

To verify the occurrence of a logarithmic term for the vacancy problem, 
we consider the following example.
Consider two dimer nets of $N^2-1$ sites each,
where $N\geq 3$ is an odd integer so that the nets admit dimer coverings.
While the two nets have different geometries, one a rectangular net of size 
$(N+1)\times (N-1)$ and one a square net of size $N\times N$ 
with one boundary odd site removed, they have the same area and perimeter. 
Any difference in the evaluationsof (\ref{dimer6}) and (\ref{dimer7}) would occur 
in ${\bar c}_3$ and ${\bar c}_3'$ and higher order terms.
Now from (\ref{dimer6}) and (\ref{dimer7}) we obtain
\bea
\ln Z_{\{(N+1)\times(N-1)\}}(1,1) &=& N^2 \Bigg(\frac G \pi\Bigg)
+ 2N {\bar c}_1 -\ln (1+\sqrt 2) +\frac 3 4 \ln 2 + \frac \pi {24}\nonumber \\
&& + \sum_{n=1}^\infty \ln \Big[1+e^{-(2n-1)\pi}\Big] +o(1)\nonumber\\
\ln Z_{\{N\times N-1\}}(1,1) &=&( N^2+1) \ \Bigg(\frac G \pi\Bigg)
+ 2N {\bar c}_1-\ln (1+\sqrt 2) + \frac 7 4 \ln 2 \nonumber \\
&& - \frac \pi {12} -\frac 1 2 \ln N+\sum_{n=1}^\infty \ln \Big[1-e^{-2n\pi}\Big] +o(1). 
\label{twoZ}
\eea
Defining the ratio
\be
R(N)\equiv \frac{Z_{\{(N+1)\times(N-1)\}}(1,1)} {Z_{\{N\times N-1\}}(1,1)}
\ee
and using (\ref{twoZ}), we find the large $N$ behavior
\be
R(N) \to C \sqrt N,\hskip .5cm N \to \infty, \label{sqrt}
\ee
where
\bea
C =\lim_{N\to\infty}\frac {R(N)} {\sqrt N}&=& \frac { e^{\pi/8 - G/\pi}} 2 
\prod_{n=1}^\infty\Bigg(\frac {1+e^{-(2n-1)\pi}} {1-e^{-2n\pi}} \Bigg). \nonumber \\
&=& 0.\ 578\ 250\ldots\ . \label{ratio}
\eea
As a numerical check, we have computed the value of ${R(N)}/ {\sqrt N} $ for $N= 3$ to $2251$ 
using (\ref{dimergd}) and (\ref{dimer2}). 
For $N=9$, for example, one has
\bea
Z_{\{10\times 8\}}(1,1) &=& 1\ 031\ 151\ 241\ =\ (89)\times (11\ 585\ 969) \nonumber \\
Z_{\{9\times 9-1\}}(1,1)&=& \ \ \ 557\ 568\ 000\ =\
2^{12}\times 3^2\times 5^3\times (11)^2 \nonumber \\
R(9)/ {\sqrt 9} &=& 0.616\ 457 \ldots\ .
\eea
Results plotted in Fig. 3 confirm the large $N$ limit of $C$ given by (\ref{ratio}) 
as well as the occurrence of the logarithmic correction in the vacancy problem.
We remark that a similar result obtained by Kenyon \cite{kenyon} involving
the occurrence of vacancy sites in the middle of a rectilinear net gives the ratio
\be
R(N) \ \sim \ N^{3/4},\hskip .5cm N \to \infty,
\ee
and thus a logarithmic correction $-{3\over 4} \ln N$ in the finite-size expansion.

\section{Summary and discussions}

We have used the Temperley-Kenyon-Propp-Wilson bijection between spanning trees 
on a lattice with free boundaries and dimer configurations on a related lattice
with a boundary vacancy to establish the independence of
the dimer generating function on the location of the vacancy. 
The equivalence is stated in Proposition (3).
The generating function for close-packed dimers on a lattice with a single boundary vacancy 
is next computed, and compared with that of the known results for dimers without vacancies. 
It is found that the vacancy introduces a logarithmic correction in the large lattice expansion.
A concrete example exhibiting this correction for an ${ MN} =$ odd net with a vacancy 
as compared to an ${ MN} =$ even net without vacancies is given.
\medskip

To ascertain whether the logarithmic correction is due to the defect of a vacancy, 
or due to the oddness of the net size, one needs to compare
expansions for two nets (of the same even-even lattice), one with two boundary vacancies
and one without vacancies.
While this problem can be formulated as the evaluation of the inverse of a matrix 
\cite{kenyon00} in the Kasteleyn formulation,
we argue that since the correction in question is that of the physical free energy
of a dimer system, on physical ground one expects the correction to be additive 
for vacancies located sufficiently far apart.
This would imply the logarithmic correction to be a ``local" property
due to the occurrence of vacancies.
\medskip

We have also found that in the language of the conformal field theory 
the central charge for the vacancy problem is $c=-2$ as compared to the value of $c=1$ 
for the dimer solution without vacancies. 
Furthermore, the $\sqrt N$ ratio (\ref{sqrt}) implies the existence of a boundary operator 
with scaling dimension 1/2, a value which does not appear in the standard Kac classification 
of operators at central charge $-2$.\footnote{We are indebted to the referee for this remark.} 
The extraction of the central charge should be viewed with caution, however,
since the dimer systems do not exhibit critical points.
\medskip

We have also carried out finite-size analyses (details of which to be given
elsewhere) for spanning trees on simple-quartic nets with other, including the toroidal, 
cylindrical, M\"obius, and Klein bottle, boundary conditions.
It is found that the logarithmic correction reported in this paper arises only
in the case of free boundaries.
This is consistent to the fact that the formulation of the extended Temperley bijection 
as presented in this paper is a property that is unique to graphs with free boundaries.

\section*{Acknowledgment}
Work has been supported in part by the National Science Council of the
Republic of China under Grant No. NSC 90-2112-M-032-003 and the
National Science Foundation Grant DMR-9980440.
We thank W. T. Lu for a useful discussion and R. Kenyon for calling our attention 
to \cite{kenyon}.
We also thank the referee for suggesting to us the formulation of Proposition 3.1.2.

\newpage
\vskip 1cm
\begin{center}
{\bf Figure captions}
\end{center}
\vskip 1cm
Fig. 1. The construction of dimer lattices $G_D$ from a
$3\times 3$ spanning tree lattice $G$.
Solid circles denote odd sites and open circles denote the odd site 
that has been removed in $G_D$.
(a) A spanning tree lattice $G$. 
(b) A dimer lattice $G_D$ constructed from $G$ with one corner site removed. 
(c) A dimer lattice $G_D$ constructed from $G$ with one odd site on the boundary removed.
\vskip 1cm

Fig. 2. The bijection between spanning trees on $G$ and dimer configurations of $G_D$.
(a) A spanning tree configuration on $G$. 
(b) The corresponding dimer configuration on the dimer lattice $G_D$ of Fig. 1(b). 
(c) The corresponding dimer configuration on the dimer lattice $G_D$ of Fig. 1(c).
\vskip 1cm

Fig. 3. The enumeration of $R(N)/\sqrt N$ for $N=$ 3 to 2251.
The dashed line indicates the value $C$ given by (\ref{ratio}) in the large $N$ limit.
\vskip 1cm
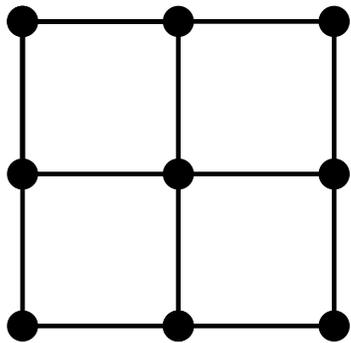
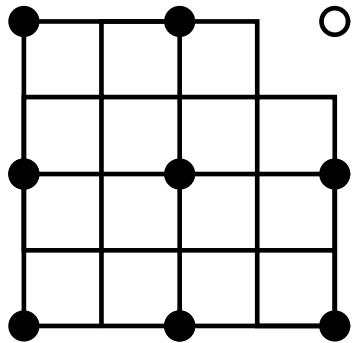
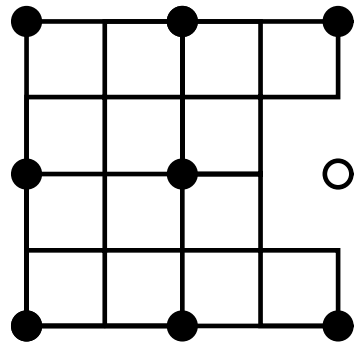
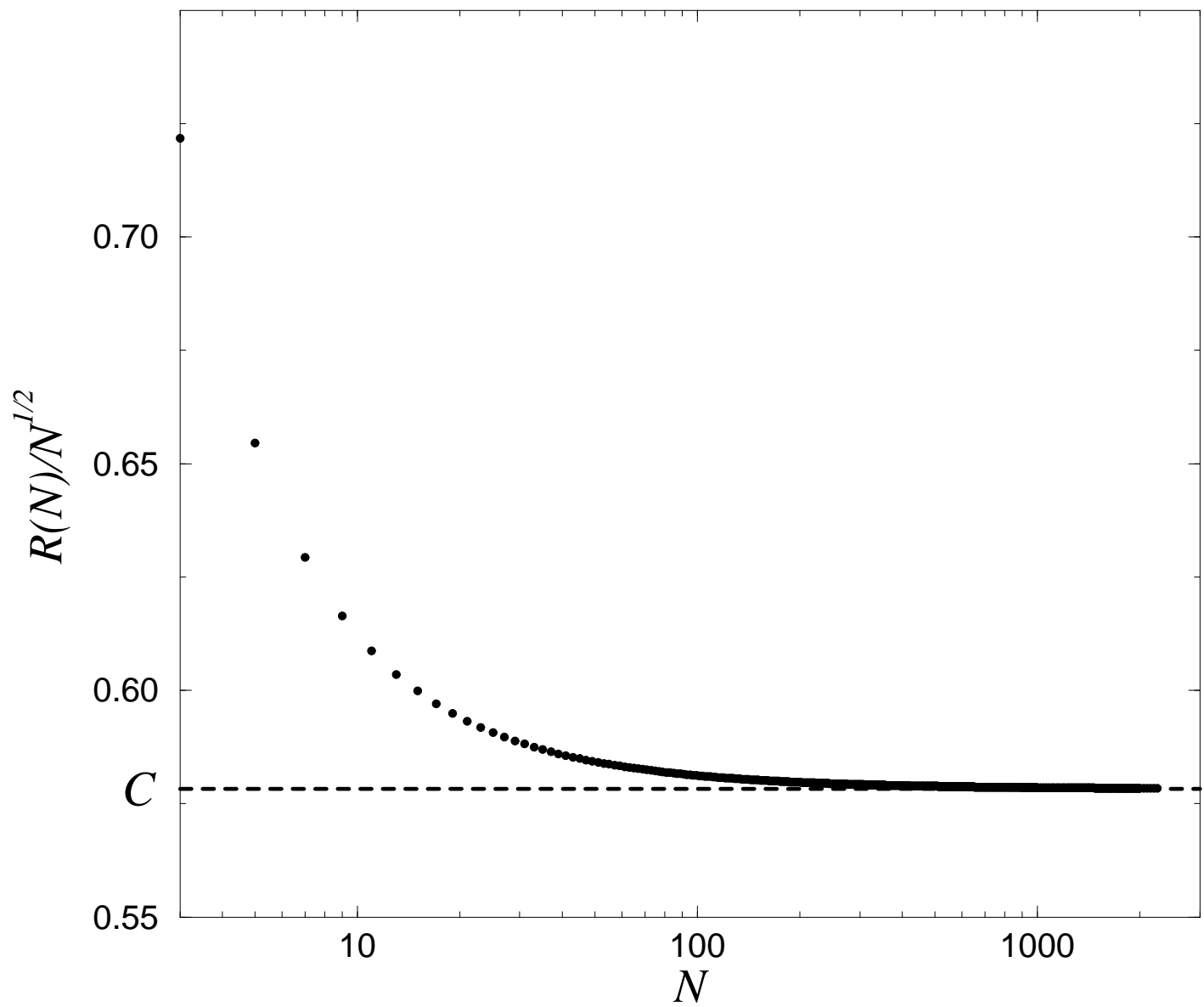
\end{document}